\documentclass[a4paper, 10pt, conference, compsocconf]{IEEEtran}

\usepackage[pdftex]{graphicx}
\usepackage{bibspacing}
\usepackage{cite}
\usepackage{multirow}
\usepackage{xcolor}
\usepackage{mdwmath}
\usepackage{mdwtab}

\usepackage[caption=false,font=footnotesize]{subfig}

\makeatletter
\newcommand*\titleheader[1]{\gdef\@titleheader{#1}}
\AtBeginDocument{%
	\let\st@red@title\@title
	\def\@title{%
		\bgroup\normalfont\large\centering\@titleheader\par\egroup
		\vskip1.5em\st@red@title}
}
\makeatother

\title{Optimization of Quality of Experience for Video Traffic}
\titleheader{\textcolor{blue}{Published in: 2015 22nd International Conference on Telecommunications (ICT). Date Added to IEEE Xplore: 18 June 2015. Date of Conference: 27-29 April 2015\\
\textcopyright 2015 IEEE.  Personal use of this material is permitted.  Permission from IEEE must be obtained for all other uses, in any current or future media, including reprinting/republishing this material for advertising or promotional purposes, creating new collective works, for resale or redistribution to servers or lists, or reuse of any copyrighted component of this work in other works.\\
DOI: 10.1109/ICT.2015.7124720}}

\author{\IEEEauthorblockN{Qahhar Muhammad Qadir$^1$, Alexander A. Kist$^1$ and Zhongwei Zhang$^2$}
	\IEEEauthorblockA{$^1$ School of Mechanical and Electrical Engineering\\
		\{safeen.qadir, kist\}@ieee.org\\
		$^2$ School of Agricultural, Computational and Environmental Sciences\\
		zhongwei.zhang@usq.edu.au\\
		University of Southern Queensland, Australia\\
}}

\begin{document}

\maketitle

\thispagestyle{plain}
\pagestyle{plain}
%
\title{Optimization of Quality of Experience for Video Traffic}


\author{\IEEEauthorblockN{Qahhar Muhammad Qadir$^1$, Alexander A. Kist$^1$ and Zhongwei Zhang$^2$}
\IEEEauthorblockA{$^1$ School of Mechanical and Electrical Engineering\\
\{safeen.qadir, kist\}@ieee.org\\
$^2$ School of Agricultural, Computational and Environmental Sciences\\
zhongwei.zhang@usq.edu.au\\
University of Southern Queensland, Australia\\
}}

\maketitle

\begin{abstract}
The rapid shift toward video on-demand and real time information systems has affected mobile as well as wired networks. The research community has placed a strong focus on optimizing the Quality of Experience (QoE) of video traffic, mainly because video is popular among Internet users. Techniques have been proposed in different directions towards improvement of the perception of video users. This paper investigates the performance of a novel cross-layer architecture for optimizing the QoE of video traffic. The proposed architecture is compared to two other architectures; \emph{non-adaptive} and \emph{adaptive}. For the former, video traffic is sent without adaptation, whereas for the later video sources adapt their transmission rate. Both are compared in terms of the mean opinion score of video sessions, number of sessions, delay, packet drop ratio, jitter and utilization. The results from extensive simulations show that the proposed architecture outperforms the non-adaptive and adaptive architectures for video traffic. 
\end{abstract}


\IEEEpeerreviewmaketitle

\section{Introduction}

The rapid shift toward video on-demand and real time information systems has affected the mobile as well as wired networks. For example, industries have started to secure their system and doctors are seeing their patients remotely using video-based systems. This increase in video based applications has contributed substantially to the growth of the Internet traffic in general and mobile data traffic more specifically. In 2013, the global mobile traffic grew by 81\% and video is expected to form 69.1\% by 2018 \cite{Cisco2014}. This is due to the fact that video has much higher bit rates compared to other contents.

The research community has recently focused on optimizing the Quality of Experience (QoE) of video traffic mainly because it has received high popularity among Internet users. Techniques have been proposed in different directions towards improvement of the perception of video users. Although, these techniques have enabled the Internet to provide some level of guaranteed QoS to realtime traffic, some consider the Internet to be broken in terms of handling traffic \cite{Roberts2009}. 

\begin{figure}[!t]
\centering
\includegraphics[width=\linewidth]{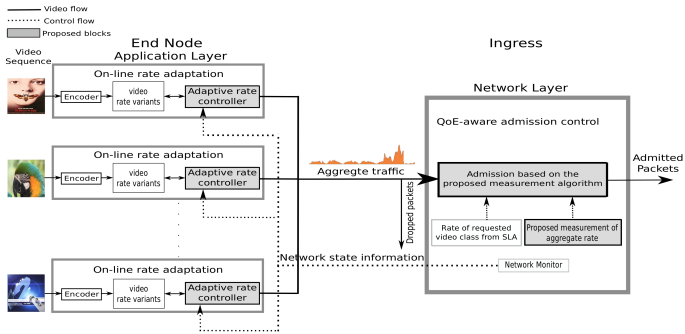}
\caption{QoE-aware cross-layer architecture for video traffic over the Internet}
\label{Proposed_Framework_paper3_conf}
\end{figure}

QoE extended the scope of expectation further wide to include the involvement of other layers in addition to network layer. In previous work, we proposed a QoE-aware cross-layer architecture for video traffic over the Internet \cite{Qadir2014}. It addressed QoE degradation in a bottleneck network by adapting video application rates and control of video session admittance. Figure \ref{Proposed_Framework_paper3_conf} shows the cross-layer architecture in which the proposed blocks are highlighted. Video sources perform rate adaptation at the application layer and the network layer controls acceptance to the ISP network through a QoE-aware admission control. Parameters from relevant layers in the application and network layers are employed by the proposed architecture. The key parameters considered are the instantaneous arrival rate of each video session and the rate of requested video session from the application layer. On the network layer, the link capacity and number of current video sessions are taken into account. For more detail about the implementation of the architecture, we refer interested readers to \cite{Qadir2014}.

In this paper, we investigate the performance of the proposed architecture (\emph{cross architecture}) in terms of the Mean Opinion Score (MOS) of video sessions, number of session, delay, packet drop ratio, jitter and utilization. It is compared with two other architectures; \emph{non-adaptive} in which video traffic sent without adaptation and \emph{adaptive} in which video sources adapting their transmission rate.

The remainder of the paper is organized as follows. Related work is reviewed in Section \ref{sec:literature}. Section \ref{sec:evaluationenvironment} explains the evaluation environment. The results are presented and discussed in Section \ref{sec:resultsanddiscussion}. The paper is concluded in Section \ref{sec:conclution}.

\section{Related Work}
\label{sec:literature}

Algorithms have been proposed to enhance QoE of video sessions through adapting transmission rate. \cite{Piamrat2009} presented a QoE-aware tool based on which rate is adapted dynamically for multicast in wireless LAN. A model has been developed by \cite{Khan2010} to optimize QoE and wireless network utilization through SBR adaptation. In \cite{Sieber2013}, a QoE-based adaptation algorithm for H.264/SVC Dynamic Adaptive Streaming over HTTP (DASH) streaming is introduced.

QoE optimization has also been proposed through cross-layer design. The authors of \cite{Zhao2014} introduced a QoE-aware cross-layer optimization to allocate the wireless resource for each DASH user. In \cite{Ai2012}, the authors set up an algorithm to allocate resources for video streaming in high speed downlink access which considers QoE, and synchronously guarantees fairness to users. An application/MAC/physical cross-layer architecture that enables optimizing perceptual quality for delay-constrained scalable video transmission has been proposed by \cite{Khalek2012}. A cross-layer scheme for optimizing resource allocation and user perceived quality of video applications based on QoE prediction model that maps between object parameters and subject perceived quality has been proposed by \cite{Ju2012}. \cite{Fiedler2009} has enhanced automatic feedback of end-to-end QoE to the service level management for better service quality and resource utilization by presenting a QoE-based cross-layer design of mobile video systems. \cite{Politis2012} incorporates a rate adaptation scheme and the IEEE 802.21 media independent handover framework to propose a QoE-driven seamless handoff scheme. \cite{Khan2006} introduced an application-driven objective function that jointly optimizes the application layer, data-link layer and physical layer of the wireless protocol stack for video streaming. \cite{Thakolsri2009} extends \cite{Khan2006}'s work from application-driven to a QoE-based cross-layer design framework for high speed downlink packet access to maximize user satisfaction. \cite{Latre2011} shows several techniques to optimize QoE in multimedia network in terms of the number of admitted sessions and video quality. Traffic adaptation, admission control and rate adaptation are proposed within an automatic management layer. The average number of satisfied users was maximized through a QoE-aware scheduling framework by sending a single bit feedback to indicate the satisfaction level \cite{Lee2014}.

Similar to the discussed literature, the \emph{cross architecture} is based on joint functionalities across layers of TCP/IP, however it is different as it integrates the rate adaptation capability of video application with a QoE-aware admission control.

\section{Evaluation environment}
\label{sec:evaluationenvironment}
For the purpose of evaluation, the \emph{cross architecture} was compared to each of the \emph{non-adaptive} and \emph{adaptive} architectures. Mean MOS, number of admitted sessions, packet drop ratio, mean delay, mean jitter and utilization were calculated as performance metrics of each architecture.

We encoded the same video sequence shown in Table \ref{tab:parameter_settings} with fixed video quantizer scale for \emph{non-adaptive} architecture and variable scales for \emph{adaptive} and \emph{cross architecture} using ffmpeg \cite{ffmpeg2004}. The details of coding parameters are recorded in the table. Source video trace files were inserted into the bottleneck link and decoded videos were generated from simulation dump files using tools provided by \cite{Lie2008} and NS-2 \cite{ns2}. We used Evalvid-RA framework \cite{Lie2008} for video encoding, decoding and interfacing with the ns-2 network. It comes with a set of tools which can be used to generate video trace files, evaluate video quality by means of PSNR and MOS metrics. The Evalvid-RA MOS metric is based on \cite{Gross2004} which computes the average quality of each video within a scale of 1 to 5 when 1 represents bad and 5 excellent quality.

A dumbbell topology with a 7Mbps bottleneck link was used to study the performance of each architecture. New sessions were requested randomly within every second of the simulation time and were accepted if the bandwidth was sufficient. In all three scenarios, in addition to 24 VBR sources, 24 FTP sources were simulated to create background traffics. Thus, maximum of 24 video sources competed for the bandwidth of the bottleneck link throughout the simulation which run for 500 seconds. The settings of the network parameters are shown in Table \ref{tab:parameter_settings}. Other parameters were kept fixed for all three architectures. 

\begin{table}[!th]
\centering
\caption{Simulation parameters}
\label{tab:parameter_settings}
\begin{tabular}{ |l|l|l| }
\hline
 & Parameter & Value \\ \hline
\multirow{5}{*}{Video content} & Video sequence & Grandma \\
 & Description & Woman speaking at low motion \\
 & Frame size & QCIF(176x144) \\ 
 & Duration(sec) & 28  \\ 
 & Number of frames & 870 \\ \hline
\multirow{4}{*}{Encoder} & Frame rate(fps) & 30 \\
 & Group of picture & 30 \\ 
 & video quantizer scale & 2 (\emph{non-adaptive}) \\
 & & 2-31 (\emph{adaptive} and \emph{cross architecture}) \\ \hline
\multirow{8}{*}{Network} & Link capacity(Mbps) & 7 \\
 & Packet size(byte) & 1052 \\
 & UDP header size(byte) & 8 \\
 & IP header size(byte) & 20 \\
 & Queue size(packet) & 2000 \\
 & Queue management & Droptail \\ 
 & Queue discipline & FIFO(First In First Out) \\ 
 & Simulation time(sec) & 500 \\ 
\hline
\end{tabular}
\end{table}

\section{Performance evaluation}
\label{sec:resultsanddiscussion}

\begin{figure}[!t]
\centering
\includegraphics[width=\linewidth]{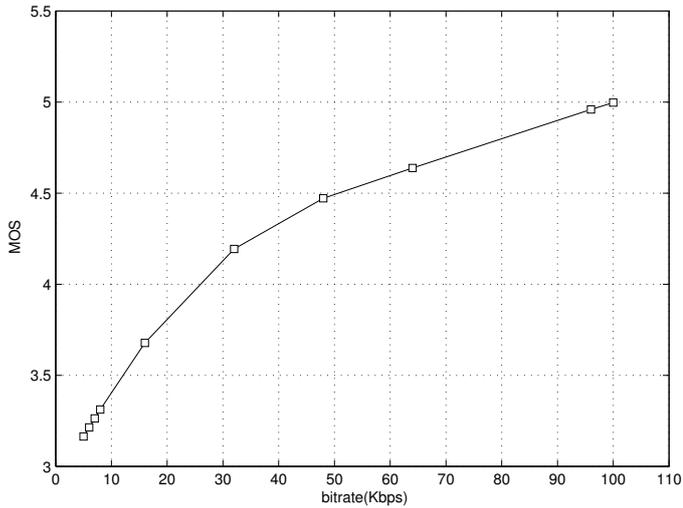}
\caption{Relationship between MOS and sender bitrate}
\label{fig:bitrate_mos}
\end{figure}

Figure \ref{fig:bitrate_mos} explains the relationship between the MOS and sender bit rate. It can be noticed that there is a logarithmic relationship between MOS and sender bitrate. A bitrate of 100Kbps or higher provides the maximum value of MOS (5) and excellent video quality for the specific video content described in Table \ref{tab:parameter_settings}. However, this relationship depends on the video content type \cite{Khan2012}. Thus, lower value of MOS and less quality is expected for medium and high content movement videos for the same bitrate \cite{Khan2012}.

\begin{figure}[!t]
\centering
\includegraphics[width=\linewidth]{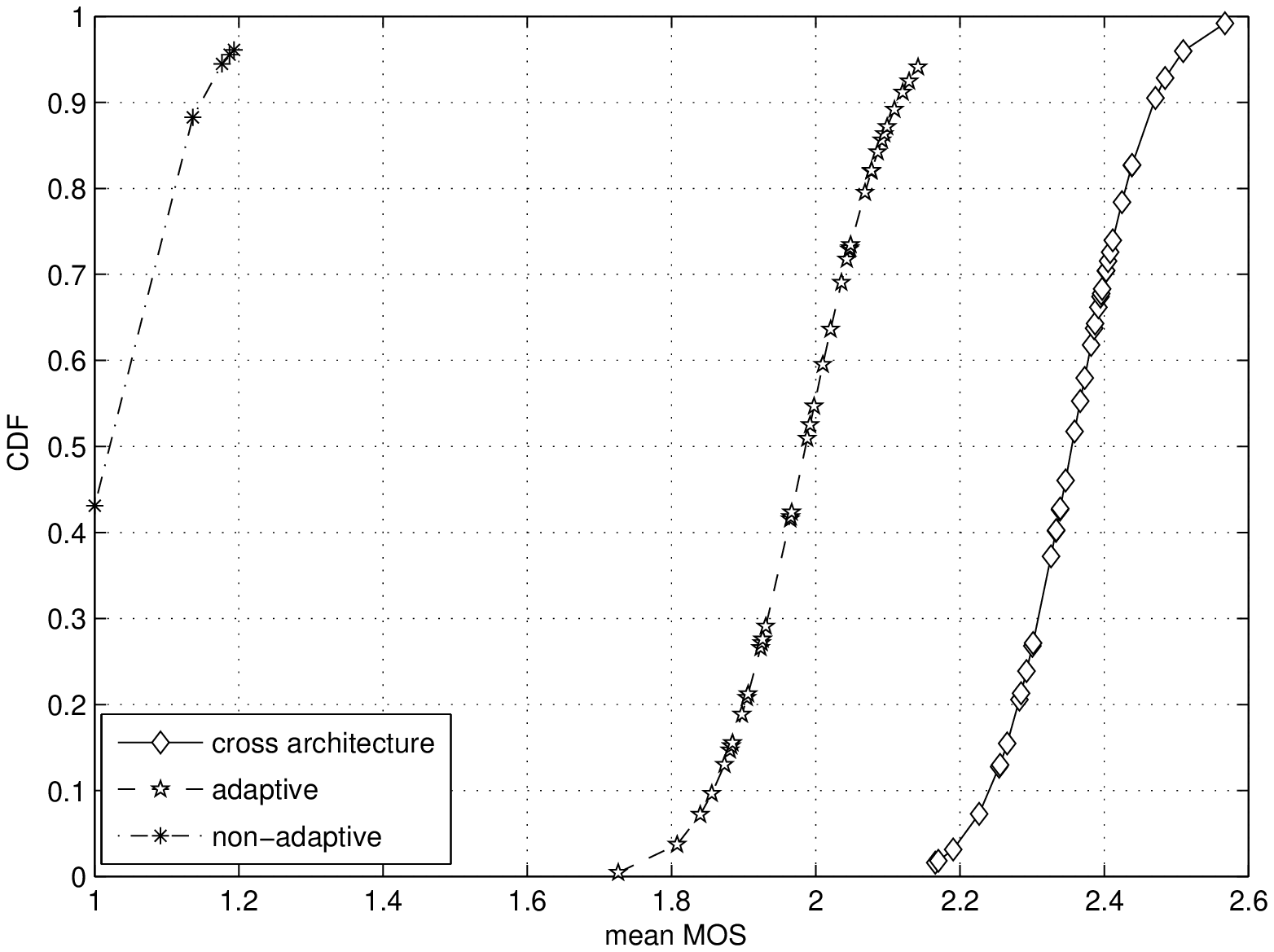}
\caption{CDF of the mean MOS of the \emph{non-adaptive}, \emph{adaptive} and \emph{cross architecture} video sessions}
\label{fig:meanMOS_all}
\end{figure}

\begin{figure}[!t]
\centering
\includegraphics[width=\linewidth]{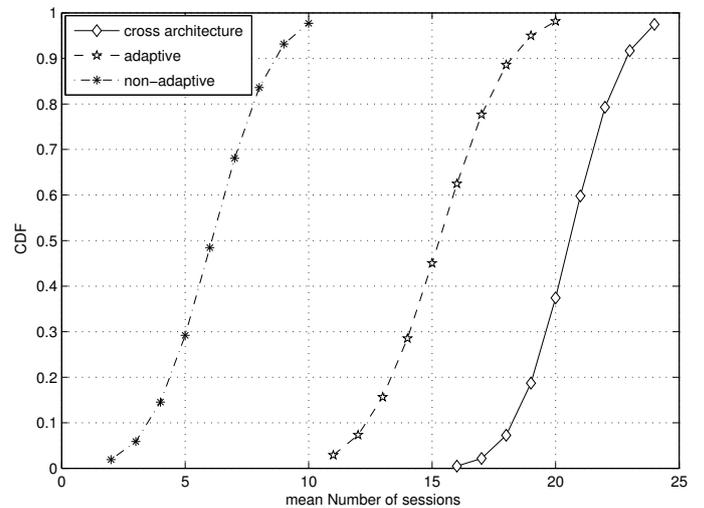}
\caption{CDF of the mean number of sessions of the \emph{non-adaptive}, \emph{adaptive} and \emph{cross architecture} video sessions}
\label{fig:meanSession_all}
\end{figure}

\begin{figure}[h]
    \centering
    \includegraphics[width=\linewidth]{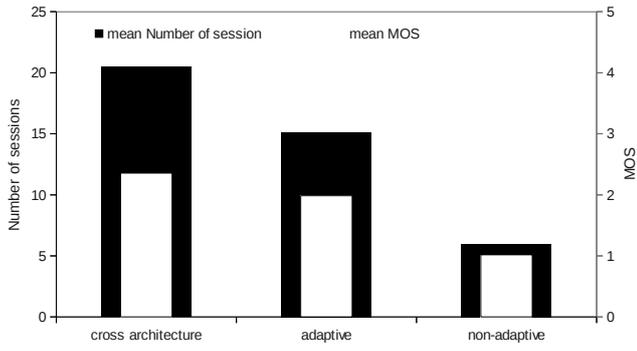}    
    \caption{mean MOS and mean number of sessions for \emph{non-adaptive}, \emph{adaptive} and \emph{cross architecture} flows}%
    \label{fig:mos-sessions_all}
\end{figure}

The Cumulative Distribution Function (CDF) of the mean MOS of accepted video sessions for each architecture is plotted in Figure \ref{fig:meanMOS_all}. \emph{adaptive} traffic attempts to adapt its sending rate according to available resources such as bandwidth and buffer. This is done at the application layer in contrast to the traditional friendly TCP self-controlling which is done at the transport layer on which the \emph{non-adaptive} architecture is based. Thus, applications that have capability of rate adaptation are more aware about what is happening in the network. Variation in the sender rate indicates the level of the quality delivered to end users. While, there is a considerable improvement of mean MOS of the \emph{adaptive} architecture (1.8) through adaptation of sender rate compared to the \emph{non-adaptive} architecture (1), a modest increase of mean MOS of the \emph{cross architecture} (2.4) over the \emph{adaptive} architecture is noticed. This slight enhancement in the video quality by the \emph{cross architecture} still can make the difference in today's huge video sessions over the Internet. Moreover, the \emph{cross architecture} accommodates higher number of sessions in addition to the modest increase of the MOS. This can be observed in Figure \ref{fig:meanSession_all}. 20 sessions with the mean MOS of 2.4 out of the total 24 video sessions are accepted by the \emph{cross architecture}, while 15 by the \emph{adaptive} and 5 by the \emph{non-adaptive} architecture. Note that we considered sessions that can be successfully decoded and played back by the receiver. i.e. there might be more sessions, but since they couldn't be decoded and played back successfully by the receiver, they were not taken into consideration. The difference of the mean MOS and number of sessions can be better compared in the bar chart \ref{fig:mos-sessions_all} which plots the CDF of the mean MOS and number of sessions together for all three architectures.

\begin{figure}[!t]
\centering
\includegraphics[width=\linewidth]{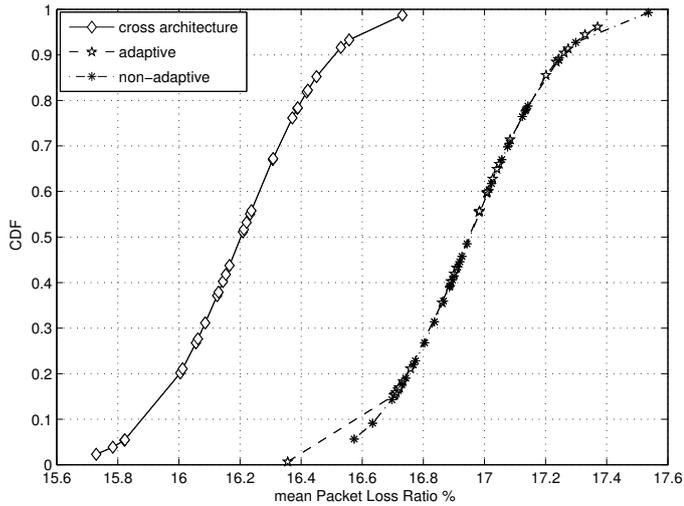}
\caption{CDF of the mean packet loss ratio of the \emph{non-adaptive}, \emph{adaptive} and \emph{cross architecture} video sessions}
\label{fig:meanPktLossRatio_all}
\end{figure}

\begin{figure}[!t]
\centering
\includegraphics[width=\linewidth]{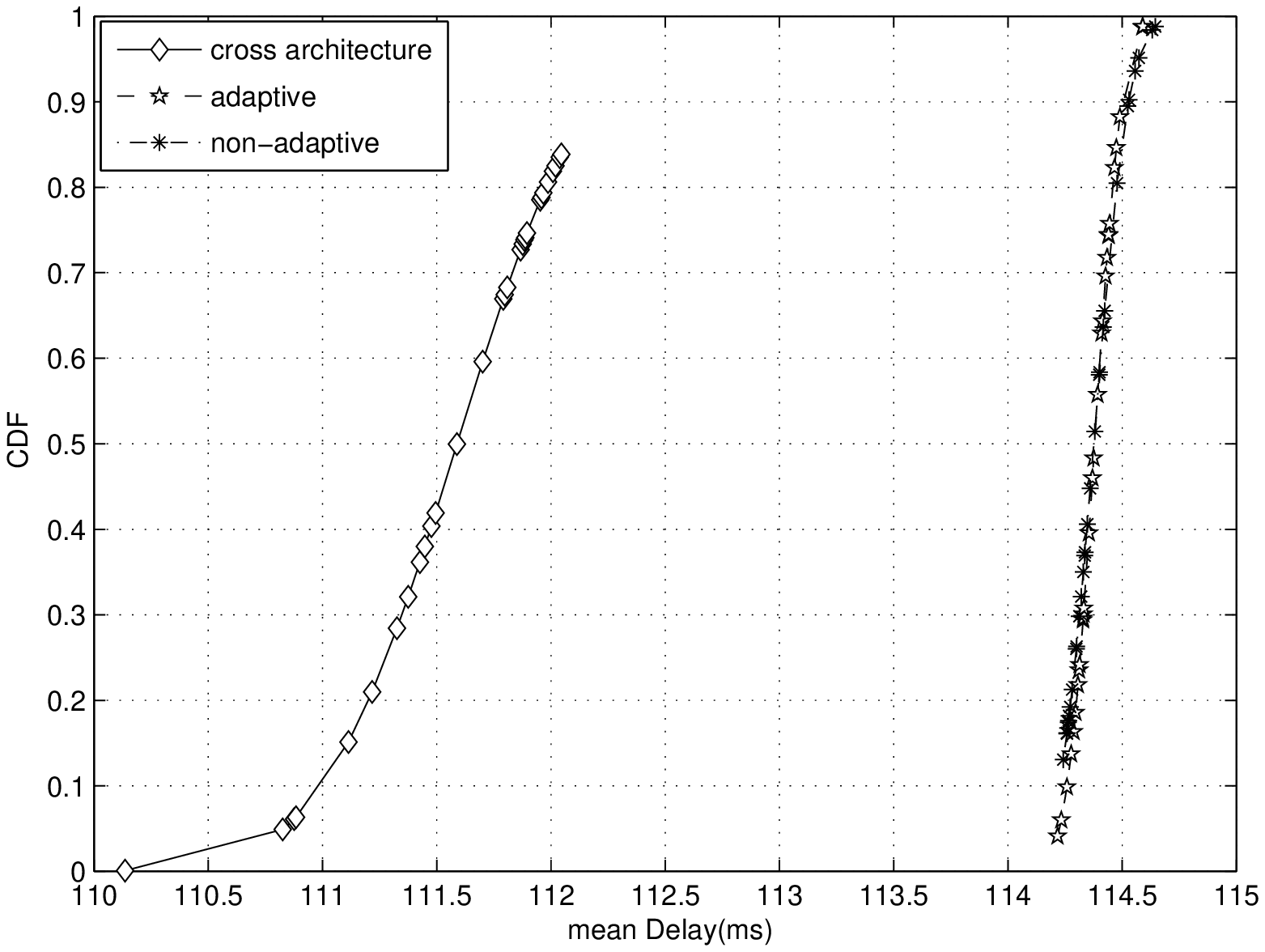}
\caption{CDF of the mean delay of the \emph{non-adaptive}, \emph{adaptive} and \emph{cross architecture} video sessions}
\label{fig:meanDelay_all}
\end{figure}

\begin{figure}[!t]
\centering
\includegraphics[width=\linewidth]{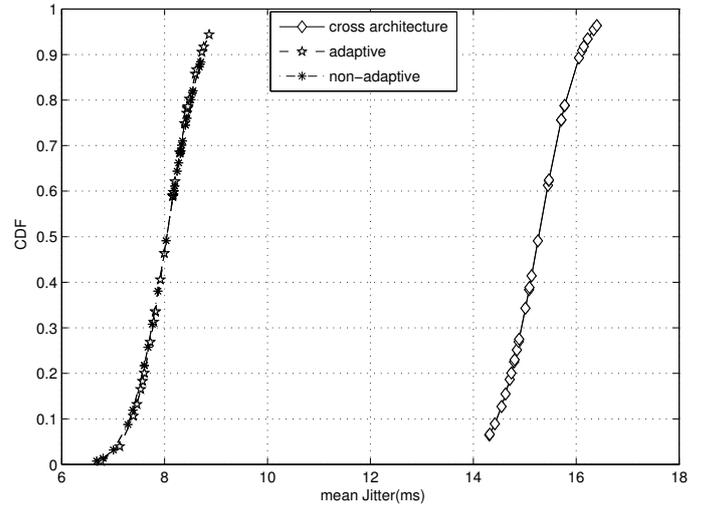}
\caption{CDF of the mean jitter of the \emph{non-adaptive}, \emph{adaptive} and \emph{cross architecture} video sessions}
\label{fig:meanJitter_all}
\end{figure}

\begin{figure}[!t]
\centering
\includegraphics[width=\linewidth]{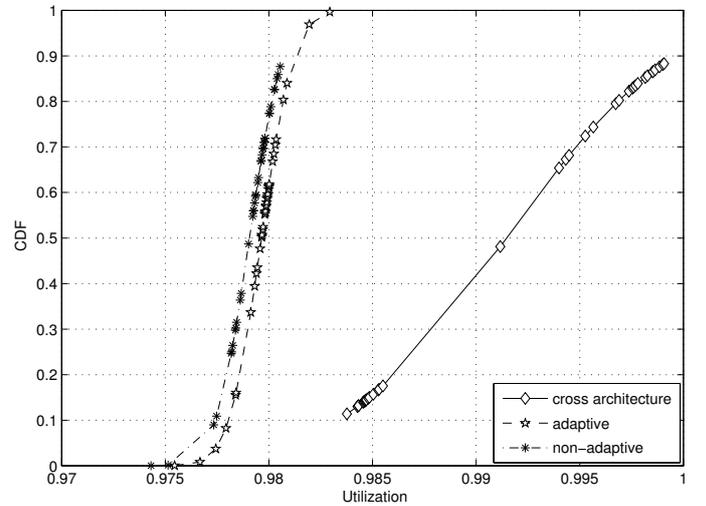}
\caption{CDF of the utilization of the \emph{non-adaptive}, \emph{adaptive} and \emph{cross architecture} video sessions}
\label{fig:utilization_all}
\end{figure}

Video streaming services are tolerance to some extend of packet loss. Thank to some characteristics of video traffic such as burstiness in the rate and error concealment in the decoder which make video to accept some tolerance of packet loss. We calculated the CDF of the mean packet drop ratio for each architecture and plotted in Figure \ref{fig:meanPktLossRatio_all}. The video traffic in the \emph{cross architecture} faced less drop compared to the \emph{non-adaptive} and \emph{adaptive} traffic.

Although, video streaming is sensitive to delay more than it is to packet loss, current massive amount of video traffic on the Internet requires less restrict technique in order to serve as maximum number as possible of user sessions with not excellent but reduced quality. The CDF of the mean delay and mean jitter for each architecture are measured and depicted in Figures \ref{fig:meanDelay_all} and \ref{fig:meanJitter_all} respectively. The video traffic in the \emph{cross architecture} experienced less delay and higher jitter compared to \emph{non-adaptive} and \emph{adaptive} architectures. The mean jitter is almost double which may be a concern for video streaming and therefore requires more attention and research.

The \emph{non-adaptive} and \emph{adaptive} architectures do not use the capacity of the bottleneck link efficiently. As mentioned earlier in this section, a few number of video sessions can be decoded and played back successfully by the receiver. However, they utilized the link considerably high in terms of the number of received bits. The utilization therefore can be seen high, although it is still lower than the utilization of the \emph{cross architecture}. Figure \ref{fig:utilization_all} shows the utilization of each architecture.

\section{Conclusion}
\label{sec:conclution}

The performance of our proposed architecture (\emph{cross architecture}) was analyzed and compared to two other architectures; \emph{non-adaptive} and \emph{adaptive}. The performance metrics taken in the study were the mean MOS of video sessions, number of sessions, packet drop ratio, mean delay, mean jitter and utilization. The extensive simulation results have shown that the \emph{cross architecture} can provide a modest improvement in the MOS, considerably a higher number of successful decoded video session, less mean delay and packet loss but a higher mean jitter. It utilizes the link more efficiently. Investigation will be conducted in the future in regards to the jitter of video traffic for the \emph{cross architecture}. Higher video resolutions will be used for the evaluation. The architecture is compared to other cross-layer architectures in further work.


\end{document}